\documentstyle[graphicx,float]{article}

\begin{document}
\title{Diffraction of entangled particles by light gratings}
\date{}
\author{Pedro Sancho \\ Centro de L\'aseres Pulsados CLPU \\ Parque Cient\'{\i}fico, 37085 Villamayor, Salamanca, Spain}
\maketitle
\begin{abstract}
We analyze the diffraction regime of the Kapitza-Dirac effect for
particles entangled in momentum. The detection patterns show
two-particle interferences. In the single-mode case we identify a
discontinuity in the set of joint detection probabilities,
associated with the disconnected character of the space of
non-separable states. For Gaussian multi-mode states we derive the
diffraction patterns, providing an example of the dependence of the
light-matter interaction on entanglement. When the particles are
identical, we can explore the relation between exchange and
entanglement effects. We find a complementary behavior between
overlapping and Schmidt's number. In particular, symmetric
entanglement can cancel the exchange effects.
\end{abstract}


\section{Introduction}

Entanglement leads to physical effects unattainable for product
states. One of the best known examples of such effects is
multi-particle interference, which has been extensively studied in
both the theoretical and experimental realms \cite{Zei,PT}. In
general these studies are within the framework of quantum optics. It
would be interesting to consider similar schemes for massive
particles. One natural candidate is the interaction of massive
particles with light gratings. In particular, the Kapitza-Dirac
effect, proposed long ago \cite{Kap,Bat}, has been experimentally
verified for atoms \cite{Pri} and electrons \cite{Ba1}.

We shall consider in this paper two-particle diffraction by light
gratings with the particles entangled in momentum, showing the
presence of two-particle interference effects absent for
factorizable states. The arrangement reflects how the light-matter
interaction is modified by entanglement. Other examples of the
dependence of the interaction on entanglement can be found in the
literature \cite{Fic,Fra}.

Similar interference effects have been already discussed for the other type of
non-product states, the (anti)symmetrized states of identical
particles \cite{Spr}. In the case of single-mode states our work
bears many mathematical resemblances to that paper, but here we
shall consider some aspects of the problem not discussed there. We
shall show the existence of a discontinuity in the set of joint detection
distributions when the two momenta are equal. This property
reflects the discontinuous nature of the space of entangled states.

For multi-mode states our approach departs from the elementary
qualitative treatment given in \cite{Spr}. Here, we consider
entangled Gaussian states and we can derive exact analytical
expressions for the joint detection patterns. These patterns,
corresponding to the sum of a large number of Gaussian terms, can
accurately be described by a single Gaussian.

The main physical novelty in the multi-mode case occurs for
identical particles. Then we can observe at work three quantum
processes (diffraction and exchange and entanglement effects) at once.
Our first finding in this scenario is the existence of a type of
complementarity between overlapping (the variable ruling the
intensity of the exchange interaction) and entanglement degree. When
one of them increases the other must decrease. To evaluate the
entanglement degree we use the Schmidt number \cite{Ebe,Fed}, which
is a measure well-suited for continuous variable problems. Another
consequence of the physical connection between particle identity and
entanglement is the cancellation of the exchange effects for
symmetric entangled states.

\section{Single-mode states}

We consider in this section single-mode states. First we describe
the arrangement and the equations describing the light-matter
interaction. In the second subsection we derive the interference
patterns in the position representation. Finally, we discuss the
singular behavior of the system when the initial momenta are equal.

\subsection{The arrangement and fundamental equations}

A sketch of the arrangement can be seen in Fig. (1). A source
generates pairs of particles in entangled states. Each particle
interacts with a light grating. The gratings, two standing light
waves formed by counter-propagating lasers with different
wavelengths, are denoted as $L$ and $R$. After the gratings we place
detectors working in coincidence to determine the joint
probabilities.

The only relevant variables in the problem are those parallel to the
grating \cite{Bat}, reducing the description of the system to a
two-dimensional one (one variable for each particle). The initial
entangled state in momentum is

\begin{equation}
|\phi _0> = \frac{1}{\sqrt 2} (|p>_L |q>_R + |q>_L |p>_R)
\end{equation}
where $p$ and $q$ denote the longitudinal momenta of the particles
and the subscripts $L$ and $R$ refer to the grating with which they
interact. Next, we must describe the interaction of the particles
with the gratings. The process is ruled by the lightshift potential
$V=V_0 \cos ^2 K x$, with $K$ the wavenumber of the laser light and
$x$ the spatial coordinate in the direction parallel to the grating
\cite{Bat}. We assume that the potential intensity is the same in
both sides but the laser wavenumbers, $K_L$ and $K_R$, differ. In
the diffraction regime of the Kapitza-Dirac effect the Raman-Nath
approximation, neglecting the kinetic part of the motion, holds
\cite{Bat}. Then, as the initial wave function for the mode $p$ in
the position representation is $\psi _0 =e^{ipx/\hbar}$, the wave
function after the interaction will be $\psi _{\tau} =e^{-iV\tau
/\hbar}\psi _0$, with $\tau$ the interaction time. The exponential
can be rewritten using the expression $e^{iz\cos \varphi}=\sum _n
J_n(z)e^{in\varphi}$, with $J_n$ the Bessel function of order $n$.
Using simple trigonometric relations we obtain $\psi _{\tau}=\sum _n
b_n e^{i(2n\hbar K+p)x/\hbar}$ with $b_n=i^n e^{-iw}J_n(-w)$ and
$w=V_0\tau /2\hbar$.

\begin{figure}[H]
\center
\includegraphics[width=9cm,height=6cm]{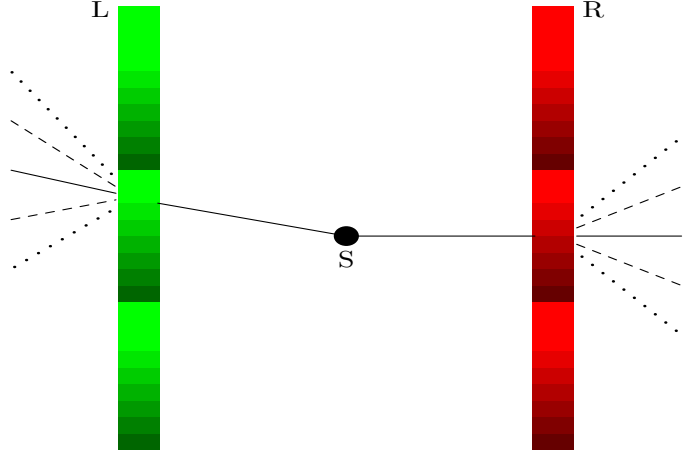}
\caption{Arrangement for two-particle Kapitza-Dirac diffraction. The
source S produces pairs of entangled particles that interact with
the light gratings L and R. The continuous, dashed and dotted lines
after the gratings correspond respectively to particles with $n=0,
\pm 1$ and $\pm 2$.}
\end{figure}

Returning to the momentum picture this evolution can be expressed as
$|p>_L \rightarrow \sum _n b_n |p+2n\hbar K_L>_L$. The final state
of the complete system is
\begin{eqnarray}
|\phi >= \frac{1}{\sqrt 2}\sum _{n,m=-\infty}^{\infty} b_nb_m (|p+2n\hbar K_L>_L |q+2m\hbar K_R>_R \nonumber \\
+|q+2n\hbar K_L>_L|p+2m\hbar K_R>_R )
\end{eqnarray}
From this expression we can derive the joint detection
probabilities. The probability of measuring the system in the state
$|p+2n\hbar K_L>_L|q+2m\hbar K_R>_R$ is
\begin{eqnarray}
P(n_p^L,m_q^R)=|_L<p+2n\hbar K_L|_R<q+2m\hbar K_R|\phi >|^2 = \frac{1}{2}|b_nb_m \delta (0)\delta (0) + \nonumber \\
\sum _{n',m'=-\infty}^{\infty} b_{n'}b_{m'} \delta (p+2n\hbar K_L-q-2n'\hbar K_L) \delta (q+2m\hbar K_R-p-2m'\hbar K_R)|^2
\end{eqnarray}
where we have used the orthogonality of the momentum states
$<p|p_*>=\delta (p-p_*)$. The above expression must be interpreted
in terms of distributions where, for instance, $b_n \delta
(0)=b_n<p+2n\hbar K_L|p+2n\hbar K_L>$. When the two conditions
\begin{equation}
2(n-n')\hbar K_L=q-p
\label{eq:cua}
\end{equation}
and
\begin{equation}
2(m-m')\hbar K_R=p-q
\label{eq:cin}
\end{equation}
simultaneously hold for some values of $n$, $m$, $n'$ and $m'$ we
have interference effects. In effect, when these conditions are not
fulfilled, $(n-n')K_L \neq (m'-m)K_R$, the detection probability
is $ P(n_p^L,m_q^R)= |b_nb_m|^2/2$ with no interference effects. In
contrast, when the conditions hold the probability becomes
\begin{equation}
P(n_p^L,m_q^R)= \frac{1}{2}|b_nb_m|^2 + \frac{1}{2}|b_{n'}b_{m'}|^2 + Re(b_n^* b_m^*b_{n'}b_{m'})
\end{equation}
The third term on the r. h. s. corresponds to the interference term
between the alternatives (i) $n$ photon absorptions in L and $m$ in
R, and (ii) $n'$ absorptions in L and $m'$ in R. The
indistinguishability of both alternatives implies that their
probability amplitudes must add, giving rise to interference
effects.

We note that, for given $K_L$ and $K_R$, only in some cases we have
interference effects. The two above conditions lead to the relation
$(n-n')K_L = (m'-m) K_R$ that can be only fulfilled when $K_L/K_R$
is a rational number as, for instance, in the case $K_L = K_R$. On
the other hand, for each pair of values $p$ and $q$ (with the
exception of, see later, $p=q$) there are always pairs of laser
wavevectors showing interference. For instance, taking $n=m=0$, $
n'=-1$ and $ m'=2$ there are interference effects for
$K_L=(q-p)/2\hbar $ and $K_L=2K_R$.

\subsection{Interference in the position representation}

In the position representation the initial entangled state is $\Psi
_0 =(\psi _L(x)\psi _R(y) + \psi _L(y) \psi _R (x))/\sqrt 2$, with
$x$ and $y$ the longitudinal spatial variables of the two particles.
After the interaction this expression becomes

\begin{equation}
\Psi _{\tau}(x,y)=\frac{1}{\sqrt 2}(\varphi _L(x)e^{ipx/\hbar}\varphi _R (y)e^{iqy/\hbar} + \varphi _L(y)e^{iqy/\hbar}\varphi _R(x)e^{ipx/\hbar})
\end{equation}
with $\varphi _L(x)=\sum _n b_n \exp (i2n K_Lx)$ and a similar
expression for $\varphi _R$. The final two-particle detection
probability, $|\Psi _{\tau}(x,y)|^2$, is
\begin{equation}
P(x,y) =P_{pro}(x,y)+Re(\varphi _L^*(x)\varphi _R(x) \varphi _R^* (y) \varphi _L(y) e^{i(q-p)(x-y)/\hbar})
\end{equation}
with $2P_{pro}=|\varphi _L(x)|^2|\varphi _R(y)|^2+|\varphi
_L(y)|^2|\varphi _R(x)|^2$ the probability of the product state
corresponding to a mixture of the initial product states $\psi
_L(x)\psi _R(y)$ and $\psi _L(y) \psi _R (x)$ with equal weights
$1/2$.

The second term on the r. h. s. of this equation represents the
two-particle interference effects, with the typical dependence on
trigonometric functions of $x-y$. At variance with momentum
interference there are no constraints between the values of
$p,q,K_L,K_R$ and the $n's$. As a matter of fact, the spatial
interference contains all the possible numbers of photon
interchanges (all the $n's$) through $\varphi$. Moreover, there is
spatial interference for any value of the ratio $K_L/K_R$. The above
properties show the different qualitative behavior of interference in
the position and momentum representations for the same initial
entangled states.

The interference effects discussed in this subsection and the
previous one are mathematically similar to those in \cite{Spr} for
identical particles. In that paper they were considered in detail
and we shall not continue here that line of argumentation, which can
be easily translated to our problem. Instead, we shall analyze
another physical aspect of the problem that was not treated there,
the existence of a discontinuity.

\subsection{Discontinuity}

Let us consider the particular case $p=q$ in Eqs. (\ref{eq:cua}) and
(\ref{eq:cin}), the conditions for the existence of interference
effects. In this case the effects only exist for $n=n'$ and $m=m'$.
However, the resultant patterns cannot be considered as genuine
two-particle interference as the necessary condition to have
interference is the existence of indistinguishable alternatives.
When $n=n'$ and $m=m'$ there is only one alternative for photon
absorption by the particles. In the case $p=q$, at difference with
the rest of values $p \neq q$, there is not two-particle
interference.

We obtain the same conclusion, in a little bit more intuitive way,
in the position representation. For instance, in the particular case
$K_L=K_R=K$, the joint detection probability becomes
\begin{equation}
P(x,y)=P_{pro}(x,y)+|\varphi _K(x)|^2 |\varphi _K(y)|^2 \cos ((q-p)(x-y))
\end{equation}
When the two momenta are equal the dependence on $x-y$,
characteristic of the spatial two-particle interference, vanishes.

This behavior can be understood in terms of the form of the
initial state $\phi _0$, which transforms for $p=q$ into
$\sqrt{2} |p>_L |p>_R $ that is a product state instead of an
entangled one. Moreover, the normalization is incorrect: the
normalized (in the single-mode sense) two-particle state is
$|p>_L|p>_R$. We have a discontinuity in the set of entangled
states. For any $p \neq q$ you have the entangled state $\phi _0$.
However, for $p=q$ there is not an entangled state.

The structure of the space of entangled states (with equal
coefficients $1/\sqrt 2$) is not $\Re ^2$ but $\Re ^2-\{p=q \}$. It
is constituted by two disconnected parts separated by the line
$p=q$. The same conclusion holds for the general states $\alpha
|p>_L|q>_R + \beta |q>_L |p>_R$, where the mathematical space is
$S^2 \otimes \{ \Re ^2 - \{ p=q \} \}$ instead of $S^2 \otimes  \Re
^2 $, with $S^2$ the Bloch sphere.

This mathematical discontinuity translates into a physical
discontinuity. If we evaluate the joint probability as the limit $q
\rightarrow p$ within the space of entangled states we have $\lim
_{q \rightarrow p} P(x,y)=2P_{pro}$. This is the joint detection
pattern for entangled momenta with $q \approx p$ (although $p \neq
q$). It doubles at any point the pattern $P_{pro}$ of a
product state. Consequently we have a discontinuity for equal values
of the momenta. In its vicinity we have joint patterns with sharply
different values. The transition between entangled and product
states is not continuous.

\section{Multi-mode states}

We consider in this section the multi-mode case. In order to deal
with analytical expressions we only consider Gaussian states, which have been extensively used in the literature (see \cite{Sou} for non-Gaussian situations). The unnormalized initial state is
\begin{equation}
\Phi _0(p,q)=e^{-p^2/Q^2}e^{-q^2/Q_*^2} e^{-pq/P^2}
\end{equation}
The coefficients $Q,Q_*$ and $P$ denote the spread of each exponential.

When $P^{-2} \neq 0$ the above state is entangled. A good measure of
the entanglement degree for continuous variable systems is the
Schmidt number \cite{Ebe,Fed}. For our problem it can be easily
evaluated analytically when $4P^4\geq Q^2Q_*^2$, giving $S=(1-
Q^2Q_*^2/4P^4)^{-1/2}$. The system is entangled when $S>1$.

The unnormalized wave function after the interaction, $\Phi _*(p,q)$,
is evaluated using two times the Fourier transform (we omit some constant coefficients)
\begin{eqnarray}
\Phi _*(p,q)=\int dx \int dy e^{-ipx/\hbar} e^{-iqy/\hbar} \sum _{n,m}
b_nb_m \int dp_0 \int dq_0 \Phi _0(p_0,q_0) \times \nonumber \\
e^{i(p_0 + 2n\hbar K_L)x/\hbar} e^{i(q_0 + 2m\hbar K_R)y/\hbar} =
\sum _{n,m} b_nb_m \int dp_0 \int dq_0 \Phi _0(p_0,q_0) \times
\nonumber \\ \delta (p-p_0-2n\hbar K_L) \delta (q-q_0-2m\hbar K_R)
=\sum _{n,m} b_n b_m \Phi _0 (p-2n\hbar K_L,q-2m\hbar K_R)
\label{eq:onc}
\end{eqnarray}
The normalized wave function is $\Phi =N\Phi _*$, with $N$ the
normalization factor. It can be explicitly evaluated when $4P^4\geq
Q^2Q_*^2$ and is given by
\begin{eqnarray}
N^{-2}=\sum _{n,m,r,s} b^*_n b_m^* b_rb_s \pi P^2 e^{-4(n^2+r^2)\hbar ^2K_L^2/Q^2} e^{-4(m^2+s^2)\hbar ^2 K_R^2/Q_*^2} \times \nonumber \\
e^{-4(mn+rs)\hbar^2 K_LK_R/P^2} e^{\alpha ^2 Q^2/8} \exp \left( \frac{P^4 \left( \beta - \frac{\alpha Q^2}{2P^2} \right)^2 }{2Q^2 \left( \frac{4P^4}{Q^2Q_*^2} -1 \right) }  \right)
\end{eqnarray}
with $\alpha =4\hbar K_L (n+r)Q^{-2}+2\hbar K_R (m+s)P^{-2}$ and $\beta =4\hbar K_R (m+s)Q_*^{-2}+2\hbar K_L (n+r)P^{-2}$.

\begin{figure}[H]
\center
\includegraphics[width=7cm,height=7cm]{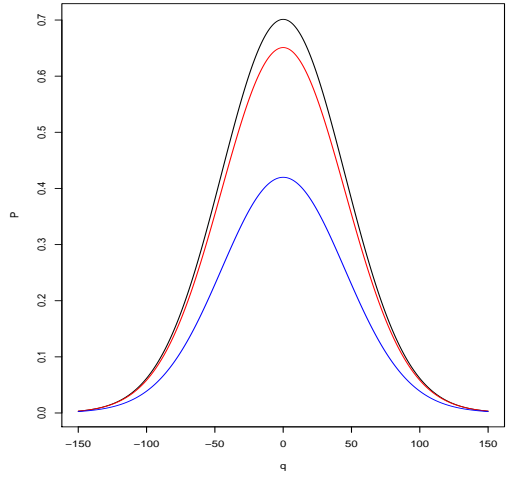}
\caption{$P(o,q)$ in arbitrary units versus $q$ in units $Q=1$. The
black, red and blue curves correspond respectively to $P=\infty,
1.1$ and $0.75$.}
\end{figure}

We give next a graphical presentation of the detection patterns
associated with these equations. We take as momentum unity $Q=1$. In
addition, we use the values $Q_*=0.9$, $\hbar K_L=0.2$ and $\hbar
K_R=0.3$. In order to see the variation of the patterns with the
entanglement degree we consider three different values of $P$:
$\infty, 1.1$ and $0.75$ that correspond to a product and two
entangled initial states. We represent in Fig. 2 the two-particle
detection pattern for $p=0$, $P(0,q)=|\Phi (0,q)|^2$. In the
evaluation we only consider the terms $n=0,\pm 1,\pm 2$.

We have that in the three cases the sum of multiple Gaussian terms
(the terms in Eq.(\ref{eq:onc})) reduces in a very good
approximation to a single effective Gaussian distribution:
$P_{eff}(0,q)=\sigma _{eff} \exp (-q^2/Q_{eff}^2)$, with the values
$\sigma _{eff}=0.7,0.65,0.42$ for $P=\infty, 1.1,0.75$, and
$Q_{eff}^2=4000$ for all the $P's$. Note that these single
distributions are not normalized (the normalization is only for the
sum over $p$ of all the $P(p,q)$).

$Q_{eff}$ is a measure of the width of the distribution. More
physically it quantifies to how many ($p=0,q$)-modes effectively
affect the diffraction. In all the cases, for product states and
for different degrees of entanglement, it is equal. On the other
hand, $\sigma _{eff}$ represents the intensity of the diffraction
process for the affected modes. In contrast with $Q_{eff}$, it
depends on the entanglement degree. In our example it decreases for
increasing Schmidt's numbers. For other values of $p$ that behavior
will be different (the total probability for all the $p's$ is the
same in the three cases).

We have also studied the form of the two-particle diffraction
patterns in the position representation. For the sake of shortness
we do not present the complete analysis here. We only signal that,
as in the momentum representation, a single Gaussian distribution
fits well the multiple Gaussian terms sum.

\section{Identical particles}

In this section we study the multi-mode case when the two particles
are identical. It is interesting because in addition to the
interference and entanglement effects we must consider the exchange
ones. The last effects are only relevant when the overlapping
between the particles is large. Then the two particles must be
diffracted by the same light grating (see Fig. 3).
\begin{figure}[H]
\center
\includegraphics[width=5cm,height=5cm]{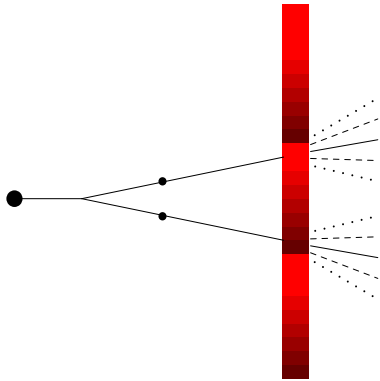}
\caption{The same as Fig. 1 but with only one light grating. The
mother particle decays into two daughter ones that interact with the
grating.}
\end{figure}

\subsection{Entanglement and overlapping}

First of all we must analyze the relation between entanglement
degree and overlapping, the two measures that determine the
intensity of the entanglement and exchange effects. The state of two
identical particles (described by the entangled state $\Phi _0$ previous to
the consideration of the identity conditions) is obtained via the
standard procedure of (anti)symmetrization:
\begin{equation}
\Phi _{ide}(p,q)=(2\pm  2\theta )^{-1/2}(\Phi _0 (p,q) \pm \Phi _0 (q,p))
\label{eq:ghi}
\end{equation}
with $\theta $ the overlapping of the two particles
\begin{equation}
\theta =\int dp \int dq \Phi _0^*(p,q) \Phi _0(q,p) = QQ_* \left( \frac{4P^4-Q^2Q_*^2}{P^4(Q^2+Q_*^2)^2-Q^4Q_*^4} \right) ^{1/2}
\end{equation}

The last expression is valid for $P^4 \geq Q^4Q_*^4/(Q^2+Q_*^2)^2$
(for instance, for the values used in the previous section). When
this condition does not hold the overlapping cannot be evaluated
analytically. The signs $+$ and $-$ in $\pm$ refer respectively to
bosons and fermions.

Through this paper we have not considered the spin (or electronic)
part of the quantum state because it is not relevant in the
diffraction dynamics. When the spin variables are taken into account
the (anti)symmetrization refers to the complete state and, for
example, the spatial wave function of fermions can be symmetrized.
By assuming that the two particles are in a symmetrized spin state,
$|s>_1|s>_2$ or $(|s>_1|-s>_2 + |-s>_1 |s>_2)/\sqrt {2}$, we do not
need to consider these cases here. With this choice the spatial part
of the state must be symmetrized for bosons and antisymmetrized for
fermions. The extension to antisymmetrized spin states is immediate.

Although there is some controversy on the characterization of
entanglement in systems of identical particles, following the
criterion in \cite{Ghi} it is clear that $\Phi _{ide}$ is entangled.
For instance, for fermions, the state is non-entangled if and only
if $\Phi _{ide}$ is obtained by antisymmetrizing a factorized state
\cite{Ghi}. Thus, for fermions (\ref{eq:ghi}) represents an
entangled state. Moreover, in this approach one clearly sees that
entanglement is associated with $\Phi _0$, the state that is
(anti)symmetrized. Then it is natural to assume that the amount of
entanglement in $\Phi _{ide}$ is the same of $\Phi _0$. As signaled
in the previous section we use the Schmidt number to quantify the
entanglement contained in $\Phi _0$.

Two interesting conclusions can be easily obtained:

(i) We fix $Q$ and $Q_*$ and vary $P$ over all the values showing
entanglement, $(QQ_*/2)^{1/2}\leq P < \infty$. When $P
\rightarrow (QQ_*/2)^{1/2}$ we have $\theta \rightarrow 0$ and $S
\rightarrow \infty$. The overlapping is very small, whereas the
entanglement tends to very high values. On the other hand, for $P
\rightarrow \infty$ these limits are $\theta \rightarrow
2QQ_*/(Q^2+Q_*^2)$ and $S \rightarrow 1$. It is simple to see that
this value of the overlapping is maximum. Now the system tends to a
factorized one whereas the overlapping is finite, reaching the
maximum value compatible with $Q$ and $Q_*$. We observe a
complementary behavior. The entanglement is maximum when the
overlapping is null and vice versa.

(ii) Symmetric entangled states represent a special case in the
relation between entanglement and overlapping. For symmetric
entanglement, $Q=Q_*$, the overlapping reaches its maximum value
$\theta =1$ independently of $P$. This is so for any type of
symmetric entangled state, $\Phi _{sym}(p,q)=\Phi _{sym} (q,p)$,
because $\int dp \int dq |\Phi _{sym}(p,q)|^2=1$ is due to the
normalization. For symmetric entangled states we have maximum
overlapping regardless of the particular entanglement degree of the
state.

In addition, symmetric states have another relevant property. When
we take $\Phi _0$ symmetric $\Phi _{ide}$ is undefined (it has the
form $0/0$) for fermions. No pair of fermions can be prepared in a
symmetric entangled state (with our choice of the spin part), a
property that can be seen as a natural extension of Pauli's
exclusion principle from product to non-factorizable states. In
effect, when the state is a product one, $\overline{\phi }(p) \phi
_*(q)$, the symmetry condition implies $\overline{\phi }= \phi _*$,
a relation that is forbidden by Pauli's exclusion principle.

On the other hand, for bosons, we have $\Phi _{ide}=\Phi
_0^{sym}(p,q)$ that corresponds to a  state without symmetrization
(in the sense of symmetrization for identical particles). The
exchange effects vanish: in the presence of symmetric entanglement
there are not exchange effects in two-boson systems. This result
confirms a previous analysis in \cite{sej}, where a similar behavior
in the position representation was found for pairs of non-entangled
identical bosons in a two-slit arrangement: in the cases where the
overlapping between the two bosons is large (it is increased by the
diffraction process to values close to unity) the two-boson
diffraction pattern is indistinguishable from that of a product
state and does not show the typical characteristics associated with
exchange effects.

\subsection{Diffraction patterns}

After analyzing the relation between $S$ and $\theta$ in the initial
state we evaluate the diffraction patterns. By similitude with Eq.
(\ref{eq:onc}) the final state is
\begin{equation}
\Phi _{ide}(p,q)={\cal N}\sum _{n,m}b_n b_m(\phi _0(p-2n\hbar K,q-2m\hbar K) \pm \phi _0(q-2n\hbar K,p-2m\hbar K))
\end{equation}
The calculation of the normalization factor ${\cal N}$ is a little bit more
involved than in the previous section. In $|\Phi _{ide}|^2$ we have
four terms, two direct and two crossed ones. The integration of each
one of them over the full momentum space gives a contribution to the squared
normalization factor. Moreover, we must take into account that the
crossed terms must be added to the direct ones in the case of
bosons, but subtracted in that of fermions. For instance, the contribution
of the first term is
\begin{eqnarray}
{\cal N}_{*}^{-2}=\sum _{n,m,r,s} b^*_n b_m^* b_rb_s \pi \xi ^2 \left(
1-\frac{\xi ^4}{P^4} \right)^{-1/2} e^{-4(n^2+r^2) \hbar ^2K^2/Q^2} \times \nonumber \\
e^{-4(m^2+s^2)\hbar ^2 K^2/Q_*^2} e^{-4(mn+rs)\hbar^2 K^2/P^2}
e^{\mu ^2 \xi ^2/4} \exp \left( \frac{\xi^2 \left( \overline{\mu }-
\frac{\mu \xi^2}{P^2} \right)^2 }{4 \left( 1- \frac{\xi^4}{P^4}
\right) }  \right)
\end{eqnarray}
where $\xi ^{-2}=Q^{-2}+Q_*^{-2}$, $\mu =2\hbar K
(2mQ_*^{-2}+2rQ^{-2}+(s+n)P^{-2})$ and $\overline{\mu} =2\hbar
K(2sQ_*^{-2}+2nQ^{-2}+(m+r)P^{-2})$. This expression is valid when
$P^4 \geq \xi ^4$. For other values of the parameters an analytical
expression cannot be obtained. It is simple to see that the two
direct (D) terms are equal, and the crossed (C) ones also. Finally
we have ${\cal N}^2 =2{\cal N}_D^2 \pm 2{\cal N}_C^2$. ${\cal N}_D$ is
given by ${\cal N}_*$ and ${\cal N}_C$ by a similar expression with
obvious modifications in the order of the coefficients $n,m,r,s$.

\begin{figure}[H]
\center
\includegraphics[width=9cm,height=7cm]{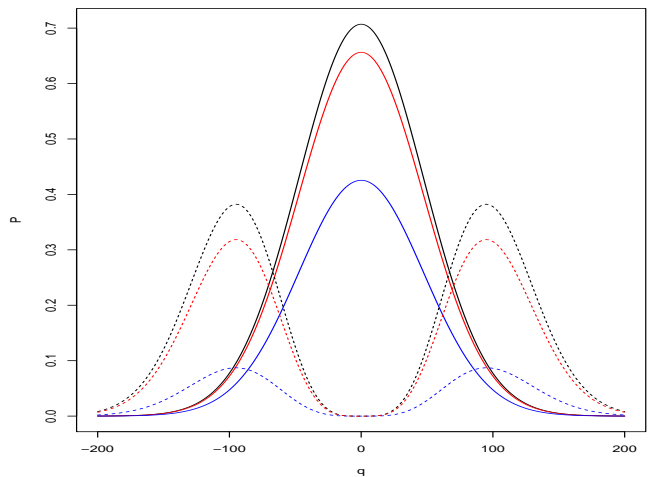}
\caption{The same as Fig. 2 with identical particles. The black, red
and blue lines correspond to the cases $P=200,1.1$ and $0.75$,
whereas continuous and dashed ones represent respectively bosons and
fermions.}
\end{figure}
Next, we represent these patterns. With the same values for $Q$ and
$Q_*$ of Fig. 2, $K=0.2$ and for the parameter $P$ the values $200$
(negligible  entanglement), $0.75$ (small overlapping) and $1.1$
(intermediate case) (see Fig. 4). We observe a clear difference between bosons
and fermions. In the first case the distribution is peaked around
the point $(0,0)$, whereas in the second this point is forbidden by
Pauli's exclusion principle. Instead, we see a valley around this
point with two peaks around. In addition to this general trend we
see that the visibility (difference between maximum and minimum
values) is strongly dependent on the entanglement degree of each
curve.

\section{Conclusions}

We have shown that entangled states are diffracted by a light
grating in a different way than product ones. This is the
essence of multi-particle interference. In an alternative view, our
results reflect the dependence of the light-matter interaction on
the separability of the state. In our proposal one can easily
visualize how the joint detection patterns change with the
entanglement degree.

The discontinuity found in the single-mode approximation provides,
in principle, a method to experimentally study the disconnected
character of the space of entangled states. It would be also
interesting to test the cancellation of exchange effects by symmetric
entanglement described above. This result, which corroborates
previous findings for non-entangled states \cite{sej}, suggests the
existence of a rich and unexplored relation between entanglement and
particle identity. For instance, the complementarity between
overlapping and Schmidt's number reflects a type of competition
between both effects.

In this paper we have restricted our considerations to thin light
gratings. For thick ones we move to the Bragg regime of the
Kapitza-Dirac effect. One expects to find similar effects to those
described here, but this must be verified by an explicit
calculation.

We have not discussed the experimental realization of the proposal.
As the  diffraction regime of the Kapitza-Dirac effect has already
been tested \cite{Pri,Ba1}, we must concentrate on the sources of
entangled particles for the arrangement. For two gratings, a
decaying particle initially at rest, seems to be a good candidate.
The single-mode case can be approximated by restricting the two
generated beams to some particular directions using collimators. For
only one grating, the photodissociation of molecules traveling
towards the standing light wave, looks to be a more adequate process
than the decay of unstable particles.

In the case of electrons a nanotip source can be interesting. The
antisymmetrization properties of such a source have been studied in
\cite{Lou}. They lead to a dip in the joint detection probability.
Although the state of electrons appears to be entangled in momentum
(Eqs. (15) and (16) in \cite{Lou}), it is not true entanglement: the
state is obtained by (anti)symmetrization of two-particle product
states (Eqs. (12) and (13) in \cite{Lou}) and then, according to the
criterion in \cite{Ghi}, the entanglement is associated with the
(anti)symmetrization procedure. However, due to the Coulomb
interaction between the electrons, there can be true entanglement in
the system. An extended analysis including that interaction would be
necessary to evaluate the entanglement and to see the viability of
an experimental test.

{\bf Acknowledgments} I acknowledge the support from Spanish Ministerio
de Ciencia e Innovaci\'on through the research project
FIS2009-09522.


\begin{thebibliography}{99}
\bibitem{Zei} M. A. Horne, A. Shimony, A. Zeilinger, Phys. Rev. Lett. 62 (1989) 2209.
\bibitem{PT} D. M. Greenberger, M. A. Horne, A. Zeilinger, Physics Today 46(8) (1993) 22.
\bibitem{Kap} P. L. Kapitza, P. A. M. Dirac, Proc. Cambridge Philos. Soc. 29 (1933) 297.
\bibitem{Bat} H. Batelaan, Contemp. Phys. 41 (2000) 369.
\bibitem{Pri} P. L. Gould, G. A. Ruff, D. E. Pritchard, Phys. Rev. Lett. 56  (1986) 827.
\bibitem{Ba1} D. L. Freimund, K. Aflatooni, H. Batelaan, Nature 413 (2001) 142.
\bibitem{Fic} Z. Ficek, R. Tana\'{s}, Phys. Rep. 372 (2002) 369.
\bibitem{Fra} H. You, S. M. Hendrickson, J. D. Franson, Phys. Rev. A 83 (2011) 023802.
\bibitem{Spr} P. Sancho, Phys. Rev. A 82 (2010) 033814.
\bibitem{Ebe} R. Grobe, K. Rz\c{a}\.{z}ewski, J. H. Eberly J. Phys. B 27 (1994) L503.
\bibitem{Fed} M. V. Fedorov, M. A. Efremov, A. E. Kazakov, K. W. Chan, C. K.  Law, J. H. Eberly, Phys. Rev. A 72 (2005) 032110.
\bibitem{Sou} R. M. Gomes, A. Salles, F. Toscano, P. H. Souto Ribeiro, S. P. Walborn, Proc. Nat. Acad. Sci. 106 (2009) 21517.
\bibitem{Ghi} G. Ghirardi, L. Marinatto, T Weber, J. Stat. Phys. 108 (2002) 49.
\bibitem{sej} P. Sancho, Eur. Phys. J. D 68 (2014) 34.
\bibitem{Lou} P. Lougovski, H. Batelaan, Phys. Rev. A 84 (2011) 023417.

\end{thebibliography}
\end{document}